%% file: main.tex
\documentclass[11pt]{article}
\usepackage[a4paper, margin=1in]{geometry}

\usepackage[utf8]{inputenc}
\usepackage{preamble, dirtytalk, textcomp, eurosym, nicefrac, dcolumn}

\usepackage{epigraph}
\title{Museums as Policy Tools: \\ The Behavioral Impact of Cultural Experiences\thanks{We are deeply indebted to Debora Barbagli and Nora Giordano, without whose generous support this experiment could not have taken place. We are grateful to the Fondazione Antico Ospedale \emph{Santa Maria della Scala} and Comune di Siena, Luca Corazzini, Pierpaolo Parrotta, Marco Stimolo, all participants to BEEN 2024 in Turin, and all participants to Soleto 2024 for their helpful comments before and after the implementation of the design. We acknowledge funds from PRIN grants P20228SXNF and 2022389MRW financed by the Italian Ministry of Research. We further acknowledge the Leverhulme International Professorship [grant number LIP-2022-001].  \newline [\href{https://www.socialscienceregistry.org/trials/14416}{Pre-registration}] 
[\href{https://drive.google.com/drive/folders/1piZa2AFSUw5IsZFwJ7N__9knxHcYbPxW?usp=drive_link}{Video Repository}]
}}

\author[s,b]{Paolo Pin}
\author[r]{Roberto Rozzi}
\author[p]{Alessandro Stringhi}

\affil[s]{Department of Economics and Statistics,
Universit\`a di Siena, Italy} 
\affil[b]{BIDSA,  Universit\`a  Bocconi, Milan, Italy} 
\affil[r]{Department of Psychology, Royal Holloway University of London, UK}
\affil[p]{Department of Management, Prague University of Economic and Business, Czech Republic}

\date{\today}

\def\sym#1{\ifmmode^{#1}\else\(^{#1}\)\fi}

\begin{document}

\maketitle

\begin{abstract}


Museums can serve as policy tools when their content is purposefully curated. We designed a framed field experiment at the \emph{Santa Maria della Scala} museum in Siena that leveraged the site’s historical role offering care and hospitality.Student visitors randomly assigned to a tour emphasizing this function later donated more to an NGO supporting refugee than those who followed a standard artistic itinerary, with effects concentrated among female participants.
These results show that thematically targeted museum experiences can measurably boost charitable behavior toward vulnerable groups, underscoring the untapped potential of cultural institutions in behavioral public policy.
\noindent \\

\vspace{0in}

\end{abstract}
\clearpage
\epigraph{
\itshape ``Cor magis tibi Sena pandit''\\
Siena opens its heart to you even more (than this gate).\\[4pt]
\upshape \small
\textit{Inscription above Porta Camollia (c.~1604), traditionally welcoming travelers and pilgrims arriving from Florence along the Via Francigena.}
}{}

\section{Introduction}\label{sec_intro}

Museums and historical sites, present in almost every city, often contain stories related to care, solidarity, and community. This makes them a promising but underused resource for shaping social attitudes and civic values. Recent discussions in the public sphere have highlighted that decisions about how museums present historical narratives can become matters of policy attention and debate.\footnote{For recent examples of the policy attention surrounding museum governance, see \href{https://www.politico.com/news/2025/08/12/smithsonian-museums-trump-review-00505838}{Politico (2025)} and the related \href{https://www.whitehouse.gov/presidential-actions/2025/03/restoring-truth-and-sanity-to-american-history/}{Executive Order from March 2025}.}
Previous research suggests that exposure to art and cultural experiences can positively influence individuals' beliefs and attitudes \citep{boyd2009origin}. This exposure is thought to evoke empathic responses, thereby enhancing prosocial behavior \citep{andries2024their}. However, focusing the experience on a given topic might improve the beneficial impact of the exposure to art on a specific societal domain. 

In this paper, we study whether cultural institutions can meaningfully influence prosocial outcomes when used in a targeted way. We compare two guided museum tours differing in how deliberately they engage with social themes. The experiment takes place at \emph{Santa Maria della Scala}, a museum in Siena that served as a hospital for a thousand years. Students in the control group participate in a standard tour focused on traditional artworks and historical artifacts unrelated to the previous role of the museum. Instead, students in the treatment group participate in a curated tour focused on the building’s former role as a medieval hospital, with emphasis on frescoes that depict care and hospitality. By subtly connecting the artistic experience to themes of social assistance, this second tour represents a more deliberate, policy-relevant use of cultural content. We measure participants' generosity through real donations to two NGOs—one operating in the field of migration, the other unrelated to those themes—and collect self-reported attitudes toward immigration in a post-tours survey. 

We find that exposure to targeted cultural experiences leads to a modest increase in prosocial behavior. Participants assigned to the treatment group donate, on average, \euro$0.25$ more than those in the control group, where the baseline donation is \euro$1.04$. Similarly, the likelihood of donating to any NGO increases by $7$ percentage points, from a baseline rate of $19\%$.
This effect is not limited to general giving: donations to the migration-focused NGO increase by \euro$0.24$, relative to a baseline of \euro$0.69$ in the control group, and the probability of donating to this NGO rises by $5$ percentage points, from a baseline of $14\%$. Importantly, these effects are not uniform across participants. The treatment appears to be concentrated among female participants. Women in the treatment group donate \euro$0.46$ more than women in the control group, while donations among men decrease by \euro$0.09$. Similarly, the probability of donating rises by $12$ percentage points for women, compared to a $2$ percentage point decrease for men. These findings suggest that thematically framed cultural experiences are effective, although they may resonate more strongly with female participants. However, despite these changes in donation patterns, we do not observe any significant shift in self-reported attitudes toward migration, as measured through survey questions administered after the tours.

Our study connects with a growing literature in behavioral economics showing that small, non-coercive interventions—so-called “nudges”—can effectively promote prosocial behavior in real-world settings. Previous field experiments have shown that subtle changes in how choices are presented, such as highlighting social norms or increasing visibility of prosocial actions, can significantly increase charitable giving, resource conservation, and cooperation \citep{frey2004social,allcott2011social, ferraro2011persistence,dellavigna2012testing,allcott2014short,andreoni2017avoiding}. These interventions often work because they activate social preferences, reduce cognitive frictions, or create implicit reputational incentives \citep{akerlof2000economics,benabou2006incentives,manna2025purely}. In this spirit, our intervention can be seen as a form of \say{cultural nudge}. However, it differs in a key aspect: it does not require additional infrastructures and investments. Guided museum tours already exist and are widely accessible; the intervention lies entirely in curating their content more purposefully. By framing the tour around themes of hospitality and care, we show that existing cultural experiences can function as low-cost tools for promoting generosity and civic attitudes. This makes our approach scalable and easy to implement in settings where cultural attractions are already in place.

In the literature, similar behavioral responses—such as increased support for refugee-related causes—are often triggered through information provision treatments \citep[see][for an extensive review of the information provision literature]{haaland2023designing}. The evidence on the effects of information provision treatments is somewhat inconclusive. Some studies do not find a significant effect of correcting people's beliefs on migration \citep{lergetporer2018natives,hopkins2019muted,barrera2020facts,alesina2023immigration, rasooly2024numbers}, and some find positive effects of such interventions \citep{grigorieff2020does,haaland2020labor, facchini2022countering}. Furthermore, information provision experiments often suffer from problems of demand effects or a lack of trust in the authority delivering the information. A commonly used solution to these problems is the treatment obfuscation \citep{haaland2023designing}. We believe our intervention represents a good alternative to information provision since our treatment is very obfuscated. Our intervention does not rely on information transmission: real world issues are never mentioned by the tour guides, and therefore, the subjects have to make connections by themselves. 

Our work is not the first in employing alternative treatments to information provision. For example, other works have used visual stimuli to improve hospitality \citep{alesina2023immigration, andries2024their}. 
Specifically, \citet{andries2024their} uses a virtual reality experience---\say{Carne y Arena}---to trigger empathy in the subjects, subsequently improving their attitudes towards immigration. Conceptually speaking, the targeted cultural experience in our design differs from the experiences used in \citet{andries2024their}, since the artistic experience in our case is less direct into triggering subjects' empathy but rather it promotes a reflection on hospitality without directly mentioning real world issues. Despite this indirect link, we still find an increase in donations towards an NGO that explicitly operates with migrants and refugees. 

The potential policy implications of our findings are supported by the literature on the effects of various forms of entertainment on individuals' behavior. Indeed, it has been shown that exposure to specific entertainment often has strong effects on attitudes or behaviors directly addressed in the medium \citep[see][for extensive reviews]{dellavigna2015economic,la2016mass}. Evidence suggests that entertainment can play both positive \citep{ferrara2012soap,kearney2015media,banerjee2019entertaining} and negative \citep{card2011family, bursztyn2023opinions, ang2023birth} roles in shaping individual attitudes. We believe that our work adds a contribution to the growing experimental literature exploring how exposure to targeted forms of entertainment can induce positive changes of attitudes \citep{greiner2007antenna, vogt2016changing, efferson2018behavioural}. 

The remainder of the paper is organized as follows. In Section~\ref{sec_design}, we present the design of the experiment in more detail, in Section~\ref{sec:results}, we present and discuss our results, in Section~\ref{sec:policy}, we provide some policy implications, and in Section~\ref{sec:conclusion}, we offer some concluding remarks. 

\section{The Experiment}\label{sec_design}

\subsection{Institutional background}
\href{https://www.santamariadellascala.com/}{\emph{Santa Maria della Scala}} is a museum located in the city center of Siena, Italy. Documented as early as 1090, it was among the earliest civic hospitals in Europe and remained in continuous operation until its closure in 1995. During the medieval period, Siena occupied a strategic position along the \emph{Via Francigena}, the pilgrimage route connecting Canterbury to Rome. \emph{Santa Maria della Scala} was originally founded to care for both local residents and pilgrims, offering food, shelter, and medical attention. Over time, the institution evolved into a modern hospital and, at the end of the 20\textsuperscript{th} century,\footnote{One of the authors of this paper was actually born there.} was repurposed as a museum that now preserves its artistic, architectural, and documentary heritage.

Like many large medieval civic hospitals (\emph{spedali}), \emph{Santa Maria della Scala} served multiple welfare functions beyond medical care. It  hosted pilgrims, offering rest, nourishment, and care at their Siena stopover on months-long journeys. For the local population, it functioned as a civic hospital with notably advanced practices for its time, including individual beds for patients and the presence of trained physicians and surgeons. The institution also played a central role in caring for abandoned children: infants left at the hospital’s foundling wheel were placed with wet nurses, later trained in a trade, and—upon reaching adulthood—granted savings and, for girls, a dowry. In addition to these core services, \emph{Santa Maria della Scala} ran soup kitchens and, by the 16th century, housed a small surgical school. Its detailed administrative records, preserved from as early as the 13th century, provide a rich historical account of its charitable mission. These layered functions make it a particularly vivid example of medieval hospitality in practice—one that remains well-suited to contemporary experimental research on themes of care, generosity, and social inclusion.

The building’s long history as a medieval hospital is still visible in its architecture and artworks, making it a particularly suitable place to explore themes like \emph{hospitality, care, and generosity}. Its layout—spread across seven floors with separate exhibition spaces—also makes it well-suited for designing \emph{randomized controlled trials}. This flexible structure allows us to offer \emph{different types of tours} to different groups while keeping the setting consistent. In our study, two areas play a central role: the \emph{Pilgrim’s Hall}, which clearly reflects the building’s original mission of assistance and hospitality, and the \emph{Fonte Gaia exhibition}, which is historically and artistically rich but unrelated to social care.

The \emph{Fonte Gaia} exhibition is the central component of the control condition in our experiment. It features the original marble sculptures from the fountain that once stood in Siena’s Piazza del Campo, created by Jacopo della Quercia in the early 15\textsuperscript{th} century. The fountain depicts biblical and allegorical scenes, including the Madonna and Child, the Virtues, and episodes from Genesis. Due to severe weathering, the sculptures were removed in the mid-19\textsuperscript{th} century and replaced with replicas carved by Tito Sarrocchi between 1858 and 1869. The originals were later transferred to \emph{Santa Maria della Scala} for conservation, along with the 19\textsuperscript{th}-century plaster casts used to produce the copies now in the piazza. While the exhibition holds significant artistic and historical value, it is unrelated to the hospital’s original mission of care and hospitality. As such, it offers an aesthetically engaging but thematically neutral setting for a standard guided tour.

In contrast, the \emph{Pilgrim’s Hall} (\emph{Sala del Pellegrinaio}) forms the core of the treatment condition. Built around the early 14th century and reaching its final structural form by the 1380s, it served as a large reception ward for pilgrims and the sick arriving at \emph{Santa Maria della Scala}. Between approximately 1439 and 1444, it was decorated with a major fresco cycle commissioned during the rectorship of Giovanni di Francesco Buzzichelli, executed by prominent artists including Lorenzo Vecchietta, Domenico di Bartolo, and Priamo della Quercia. Domenico di Bartolo’s famous panels—such as Care of the Sick, Distribution of Alms, and Reception of Pilgrims—offer realistic depictions of hospital life: physicians, friars, patients, soup distribution, orphan care, and the daily charity routines that characterized the institution’s mission.
These images act as powerful visual narratives of hospitality, care, and social support rooted in the institution’s original charitable mandate. The treatment tour is intentionally centered on this hall, focusing participants’ attention on the artwork and architecture that embody the hospital’s historic function—without ever invoking contemporary themes like migration or modern philanthropy. This symbolic framing allows us to investigate whether exposure to these motifs, in a deeply contextualized historical setting, can influence present-day attitudes and behavior.

\emph{Santa Maria della Scala} shares much with other major medieval hospitals that have since become museums—such as the Ospedale degli Innocenti in Florence, the Hôtel-Dieu in Beaune, and St Bartholomew’s in London. Like these institutions, it preserves the architecture, artworks, and historical memory of premodern social care. What sets \emph{Santa Maria della Scala} apart, and makes it especially well-suited for our experimental design, is the combination of preserved hospital spaces, thematically rich visual narratives like those in the Pilgrim’s Hall, and a modular museum layout that enables the controlled variation of visitor experiences. While some features are site-specific, the conceptual framework of our intervention—using historical environments to evoke values such as hospitality and generosity—can be adapted to other settings where cultural heritage intersects with social themes.

\subsection{Experimental design}

The experiment consists of two core components: a 45 minutes guided cultural tour within the \emph{Santa Maria della Scala} museum and a post-tour survey capturing incentivized and self-reported measures of prosociality and hospitality. All participants receive a fixed show-up fee of \euro20, paid in the form of an Amazon gift card sent via email within two weeks after their session. After completing the cultural activity with museum employees, participants are handed a printed questionnaire by the experimenters. This paper form includes both the donation task—where participants could allocate any portion of their \euro20 compensation to one or both of two non-profit organizations—and a set of survey questions. The two NGOs are the \emph{Consiglio Italiano per i Rifugiati} (CIR), which provides support to refugees and asylum seekers, and the \emph{Lega Italiana Protezione Uccelli} (LIPU), an environmental organization focused on the protection of birds and their habitats. Participants are free to donate any amount, including zero, to either or both NGOs, and donations are deducted directly from their compensation. Experimenters record all the donations and then take care of sending the total amount to the two NGOs at the end of the whole experiment. The same form also includes questions from the European Social Survey (ESS) measuring attitudes toward immigration, along with basic demographic questions and an enjoyment rating for the tour.

\paragraph{Tours in the museum.}
Participants are randomly assigned at the session level to one of two tours, which are matched in duration, structure, and delivery format, but differ in thematic focus. In both cases, professional guides employed by the museum lead the tours using scripts developed in collaboration with the research team. Importantly, students sign-up for an unspecified tour to the museum and they remain unaware of the tour they have been assigned to until the tour starts. This choice allows us to exclude self-selection into tours and to keep the randomization intact.

\emph{Non targeted artistic tour (Control).} Participants in the control condition take part in a guided tour centered on the \emph{Fonte Gaia} exhibition. The guide provides an overview of the artistic features and historical context of the fountain, with a focus on its sculptural elements, iconography, and 19th-century restoration. The tour emphasizes aesthetic and historical aspects and does not address social themes such as care, migration, or hospitality.

\emph{Targeted artistic tour (Treatment).} Participants assigned to the treatment condition visit the \emph{Pilgrim’s Hall}, where the guide focuses on the frescoes depicting the historical functions of the hospital. The commentary highlights scenes of medical care, charitable acts, and the reception of pilgrims. In particular, the tour is conducted without drawing any explicit parallels to present-day issues such as immigration, public health, or social policy. The framing remains entirely historical and descriptive, allowing us to test whether exposure to thematically relevant, but non-explicit, cultural content could influence attitudes and behavior.

Importantly, the tours are conducted in Italian by two museum employees who are professional guides. Each is responsible for designing and leading only one of the two tours. While both were aware that the tours were part of an experiment, they were not informed about the research question or the content of the post-tour questionnaire. They are also explicitly instructed to remain neutral in their delivery. Any connection between the themes of the treatment tour and concepts such as generosity or immigration is meant to emerge spontaneously in participants’ minds, without prompting or framing by the guide.\footnote{A video recording of one control and one treatment tour can be found at the  \href{https://drive.google.com/drive/folders/1piZa2AFSUw5IsZFwJ7N__9knxHcYbPxW?usp=drive_link}{following link}. Note that the recording of each type of tours is split in three files.}

\paragraph{Measures of prosociality and hospitality.}Our main outcome is based on an incentivized donation task in which participants decide how much of their \euro20 show-up fee to allocate to two non-profit organizations. This real-stakes decision allows us to measure prosocial behavior in a consequential way. Total donations across both organizations serve as a proxy for general prosociality. To isolate attitudes specifically related to hospitality, we focus on donations directed to the \emph{Consiglio Italiano per i Rifugiati} (CIR), a prominent NGO supporting refugees and asylum seekers. Because this organization operates in the domain of migration and social inclusion, contributions to it capture participants' willingness to support groups associated with hospitality-related concerns. Donations to the second NGO, the \emph{Lega Italiana Protezione Uccelli} (LIPU), which focuses on environmental protection, serve as a thematically neutral reference point.

\paragraph{Self-reported attitudes towards migration.} Subsequently, subjects had to answer seven questions from Section B of the European Social Survey (ESS) Round 7 \citep{card2005understanding}. Specifically, questions B29 to B34. These questions elicit respondents’ attitudes and beliefs toward immigration. The questions are reported in Appendix~\ref{secA_ESS_mig}. We use responses to these items to construct a standardized index capturing individual attitudes toward immigration.

\paragraph{Satisfaction questionnaire.} To ensure that our results are not simply driven by participants’ enjoyment of the tours—which could be correlated with the specific tour or guide—we ask them to rate their experience on a scale from 1 to 10. 

\paragraph{Procedures.} We recruited students from the University of Siena through HROOT. All registered native-speaking students received invitations to participate in guided tours at the \emph{Santa Maria della Scala} museum. A total of 388 Italian-speaking students took part in the experiment, of which 139 were males and 249 were females.\footnote{The distribution reflects the one of the subject pool (1248 in total, of which 785 females).} We conducted the experiment between March and April 2025 across 16 sessions, equally divided between the two conditions. While assignment to sessions was random from the students’ perspective, the actual scheduling of the tours was determined by the availability of the two professional guides, each of whom was responsible for only one type of tour.\footnote{We discuss the tour guide fixed effects and the enjoyment in Appendix~\ref{app:tables}.} The full session calendar is reported in Appendix~\ref{app:calendar}.

\subsection{Testable Hypotheses}

We design the experiment to test two primary hypotheses. First, we investigate whether exposure to targeted artistic content increases participants’ overall generosity, as measured by the total amount donated to the two NGOs. Second, we examine whether such exposure specifically enhances generosity toward causes related to hospitality, proxied by donations to an NGO operating in the field of migration.\footnote{All hypotheses and pre-analysis plans are available at the following link: \url{https://www.socialscienceregistry.org/trials/14416}. Self-reported immigration attitudes, measured through ESS questions, were included as secondary outcomes in the pre-analysis plan.}

\emph{Hypothesis 1 (H1).} Total donations are higher in the treatment group than in the control group.

\emph{Hypothesis 2 (H2).} Donations to CIR are higher in the treatment group than in the control group.

These two hypotheses capture different dimensions of generosity. H1 tests for a general increase in prosocial behavior, while H2 isolates generosity specifically directed toward a cause aligned with the treatment’s thematic content. Crucially, the two are not logically linked: total donations could increase without greater support for CIR, or CIR donations could rise at the expense of LIPU, leaving the total unchanged. For our intervention to achieve its intended effect—fostering generosity in the domain of hospitality—both hypotheses must be confirmed.

\section{Results}\label{sec:results}

This section presents the main results of the experiment. As pre-registered, we focus on the effects of the treatment on donation behavior. Additional analyses, including robustness checks and secondary outcomes, are reported in Appendix \ref{app:tables} and discussed after the main regression analysis. 

\subsection{Descriptive Evidence}

Participants exposed to the targeted artistic content exhibit higher generosity compared to those in the control group. On average, participants in the treatment group donate \euro1.29, an increase of approximately 24\% over the control group's baseline of \euro1.04. Similarly, the likelihood of making any donation rises from 19\% in the control group to 26\% under treatment—an increase of nearly 40\%.

Importantly, this increase in generosity seems to be driven by female participants. Females in the treatment group donate \euro1.65 on average, more than \euro1.19 in the control group. Their likelihood of donating rises from 19\% to 31\%. By contrast, no meaningful differences emerge among male participants, whose donation rates and amounts remain unchanged (17\% and \euro0.66 in treatment vs. 19\% and \euro0.75 in control).  

\begin{figure}[ht!]
    \centering
    \includegraphics[width=0.90\linewidth]{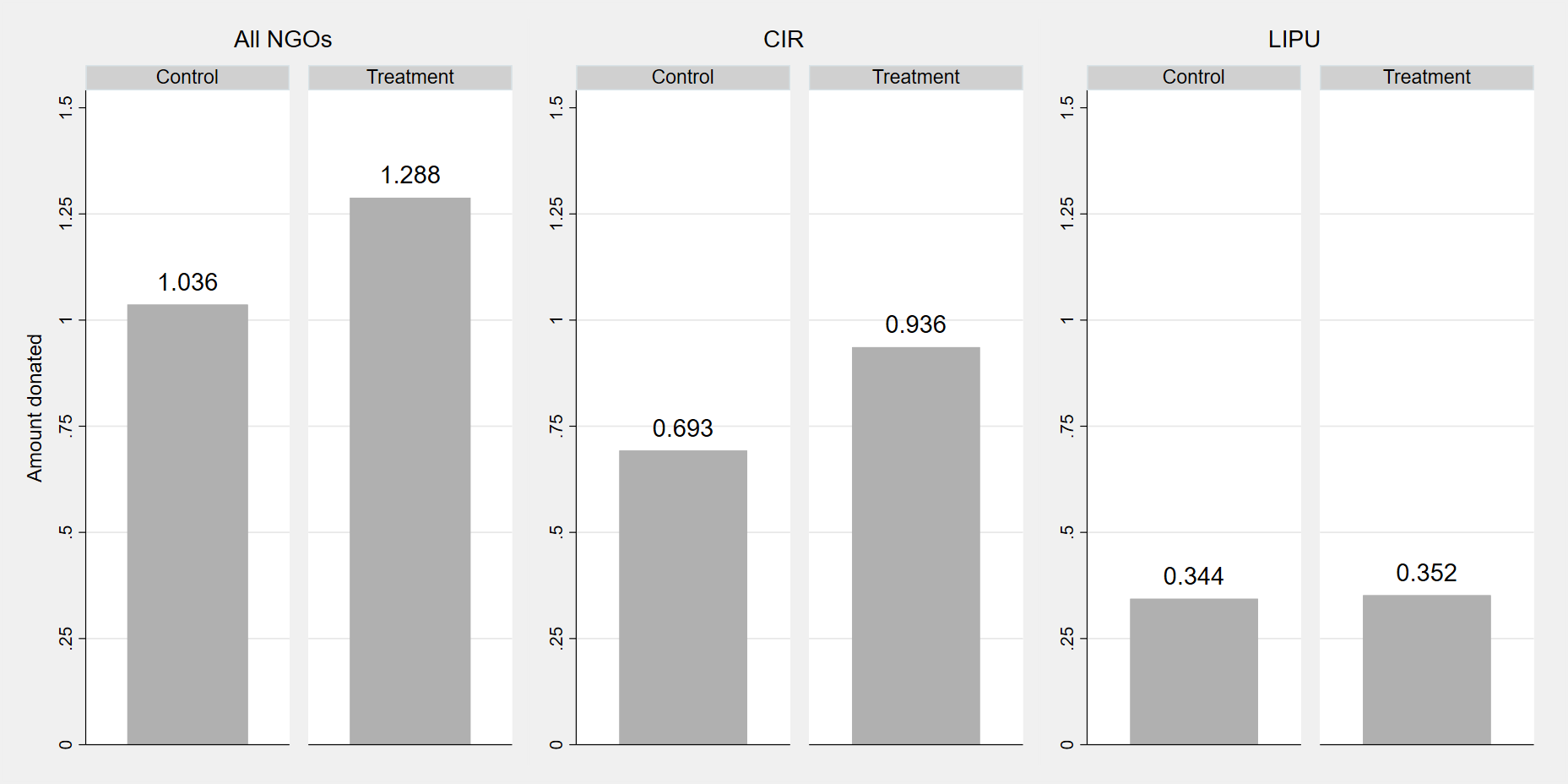}
    \caption{This figure shows the average donations to all NGOs, to CIR, and to LIPU for the Control and Treatment groups.}
    \label{fig:donations}
\end{figure}

\begin{figure}[ht!]
    \centering
    \includegraphics[width=0.90\linewidth]{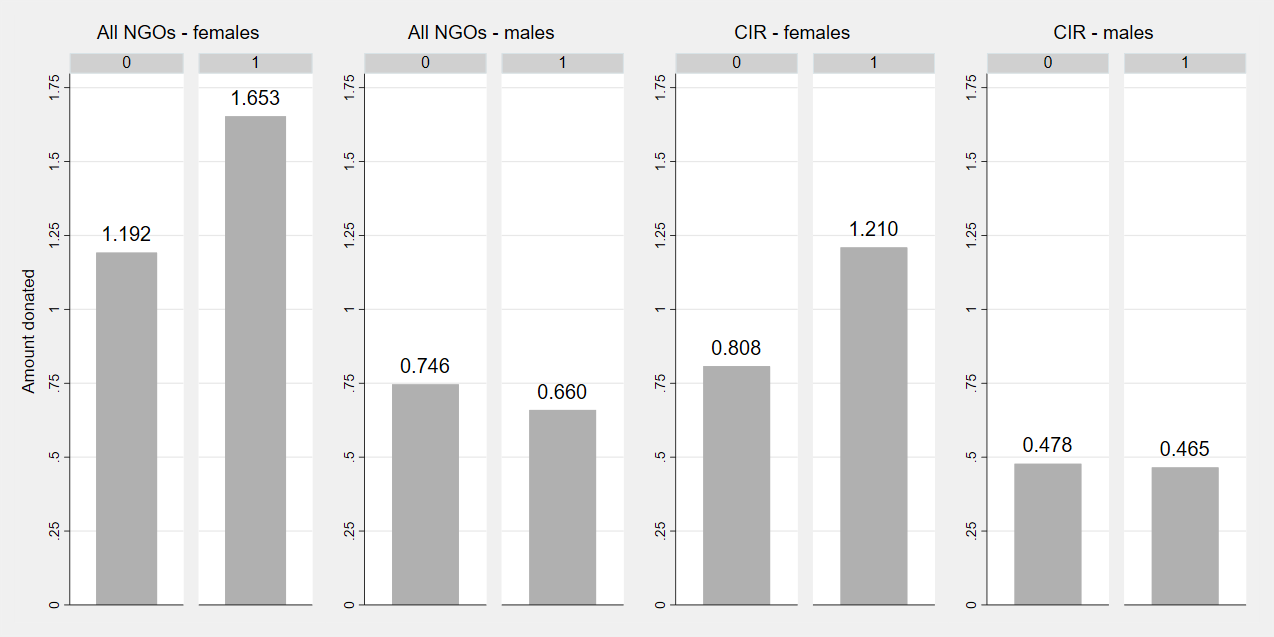}
    \caption{This figure shows average donations to all NGOs and to CIR by treatment status, separately for female and male participants.}
    \label{fig:group_gender}
\end{figure}

Looking at donations to CIR, the NGO linked to migration and hospitality, participants in the treatment group give \euro0.94 on average, compared to \euro0.69 in the control group, a 36\% increase. Again, this effect is concentrated among female participants, whose average donations to CIR rise from \euro0.81 to \euro1.21. Donations to the thematically neutral NGO (LIPU) show no significant changes, and remain consistently lower than those to CIR for both males and females. Figures~\ref{fig:donations} and~\ref{fig:group_gender} summarize the descriptive evidence discussed above. 

\subsection{Probability of Donating}

We first estimate the effect of the treatment on subjects’ donation behavior. Table~\ref{tab:probit} presents the results of a probit regression where the dependent variable indicates whether a subject donated to either NGO, and the explanatory variable of interest is a treatment dummy equal to one for subjects assigned to the treatment group.

\input{probit}

Table~\ref{tab:probit} shows that the treatment increases the probability of donating. In the full sample, assignment to the treatment raises the likelihood of donating to any NGO by $6.9$ percentage points, although the effect is only marginally significant. The effect is stronger and statistically significant for donations to CIR, where the treatment increases the probability of donating by $5.4$ percentage points, while the coefficient for LIPU is small and not statistically significant. This pattern suggests that the treatment affected the extensive margin of giving primarily toward the thematically aligned NGO. Consistent with the descriptive evidence, the effect is concentrated among female participants: for women, the treatment increases the probability of donating to any NGO by $10.1$ percentage points and the probability of donating to CIR by $7.3$ percentage points, with no detectable effect on donations to LIPU.

\subsection{Regression Analysis on Donation Amounts}

To complement the descriptive evidence, we estimate the effect of the treatment on donation amounts using a censored regression model. Our outcome of interest is the amount donated across the two NGOs, measured in euros. The key explanatory variable is a binary indicator for assignment to the treatment group. We include controls for gender, age and nationality.
Since a large share of participants opted not to donate any portion of their endowment, the distribution of donations is left-censored at zero. To account for this, we employ Tobit regressions, which are appropriate in settings where the outcome variable is censored and a substantial mass of observations lies at the censoring point. Moreover, a considerable amount of money donated in the control group comes from two outliers who donated their entire endowment and alone account for almost 20\% of the total amount. To avoid distortions caused by extreme values, we exclude three observations in the main regression analysis, including a third case in the treatment group.\footnote{See Appendix \ref{app:outliers} for a discussion.}

Results from the Tobit regressions from the full sample are reported in the first three columns of Table \ref{tab:tobit}. We find that assignment to the treatment group significantly increases total donations. In the full sample, the estimated treatment effect is positive and significant (coefficient = $1.77$, $p < 0.05$), confirming the direction observed in the descriptive analysis. The effect is also present for donations to CIR—the NGO aligned with the treatment's thematic framing—where the treatment effect is also sizable and significant (coefficient = $1.76$, $p < 0.05$). In contrast, we find no significant effect of the treatment on donations to LIPU, the thematically neutral NGO.

The results also reveal substantial gender effect. Female participants exhibit significantly higher generosity overall (coefficient = $1.92$, $p < 0.05$), and especially in donations to CIR (coefficient = $2.79$, $p < 0.01$). No significant gender effect is observed for LIPU. Age also appears to be a significant factor in the amount donated to charity. Indeed, older participants donated a larger amount of money to both NGOs, and especially to LIPU. 



\input{tobit}


To further explore the substantial gender effect suggested by the full sample regression analysis, we estimate separate Tobit models for female participants. The results for female participants, reported in the last three columns of Table \ref{tab:tobit}, confirm that only women are responsive to the treatment. Among female participants, assignment to the treatment group significantly increases the propensity to donate (coefficient = $2.48$, $p < 0.05$), with the effect driven almost entirely by donations to CIR (coefficient = $2.18$, $p < 0.05$). Results for male participants reveal no significant effects and are reported in the appendix. 

\subsection{Gender Differences}

Table~\ref{tab:ttest_gender} shows that gender differences in donation behavior emerge only among participants assigned to the treatment group. In the control group, male and female participants donate similar amounts across all outcomes, with no statistically significant differences in either donation amounts or likelihood of donating. By contrast, in the treatment group, female participants donate significantly more than male participants, both in terms of total donation amounts (\euro$1.65$ vs. \euro$0.67$, $p = 0.016$) and the probability of making a donation (30\% vs. 17\%, $p = 0.035$). This gender gap is especially pronounced for donations to CIR, where both the amount donated (\euro$1.21$ vs. \euro$0.47$, $p = 0.025$) and the likelihood of donating ($25$\% vs. $11$\%, $p = 0.021$) are significantly higher among women.

These findings reinforce the view that the observed effects of the targeted cultural experience are not uniform across genders. Instead, they appear to be driven almost entirely by female participants, complementing the findings from the main regression analysis. A plausible explanation is that, in our experiment, women are more responsive and attentive than men to the part of the treatment tour emphasizing hospitality and generosity. This interpretation is consistent with previous evidence from experimental economics and psychology showing that women are not necessarily more generous, but tend to be more responsive than men to the experimental context and, more generally, to social cues \citep{gilligan1993different,fischer20004,croson2009gender,dellavigna2013importance}.

\input{ttest_gender}

\subsection{Secondary Outcomes}

We complement the main behavioral results with two secondary analyses aimed at better understanding the mechanisms underlying the treatment effect. Specifically, we examine whether differences in tour enjoyment or changes in migration attitudes can account for the observed increase in donations.



\paragraph{Enjoyment effect.} When analyzing the satisfaction questionnaire, one can notice that enjoyment is unbalanced across the two groups. Specifically, as reported in Table \ref{tab:balance}, enjoyment is on average higher in the control than in the treatment ($p < 0.01$). This imbalance goes in the opposite direction of the donation results: participants in the control group report higher enjoyment, yet donate less than participants in the treatment group. This result allows us to rule out the hypothesis that the positive treatment effect is driven by a form of reciprocity, whereby subjects donate more because they enjoyed the tour more. 

Since each guide was associated with only one treatment condition, guide and treatment effects cannot be separately identified. However, the enjoyment imbalance works against a reciprocity-based interpretation: participants report higher enjoyment in the control condition, where donations are lower. Moreover, in diagnostic specifications reported in the appendix, enjoyment does not predict donation behavior conditional on treatment assignment and pre-treatment covariates. We therefore find no evidence that the treatment effect is driven by differential appreciation of the tour-guide experience.

\paragraph{Migration attitudes.} 

Complementing the donation results, we examine whether the targeted cultural experience shifted respondents’ self-reported migration attitudes. Table~\ref{tab:imm_attitudes} shows that the treatment has no detectable effect on participants’ migration attitudes, point estimates are small and imprecisely estimated around zero, indicating no detectable effect on the ESS-based attitude index. Participants were already highly favorable toward migrants (mean score of 0.78), potentially reducing treatment responsiveness \citep{haaland2023designing}. Importantly, this null result does not weaken the interpretation of the behavioral findings. On the contrary, it suggests that the increase in prosocial donations operates through mechanisms other than explicit belief updating—consistent with the subtle and fully obfuscated nature of the intervention, which never referenced contemporary migration issues. This pattern aligns with a broader literature showing that forms of nudges such as contextual framing, moral salience, and affective cues can shape prosocial decisions even when stated beliefs or attitudes remain unchanged \citep{de2018putting,graad2024nudges,rodemeier2025information}. 



\section{Policy Implications}\label{sec:policy}



Our findings suggest that targeted cultural experiences are best understood as tools for behavioral activation when they are linked to immediate opportunities for civic action rather than as stand-alone instruments for durable attitude change. Petitions, fundraising campaigns, or volunteer recruitment are natural examples: even if the motivational effect is short-lived, the action it induces may have persistent institutional consequences.

The narrow focus of our intervention—hospitality in a historically constrained setting—should not obscure its broader applicability: the same logic can be extended to a wide range of topics, and modern museums often possess far greater flexibility in how they frame and present content. From a policy standpoint, the required investments are minimal. Adjusting the narrative structure of existing tours or reallocating guides within the museum’s current layout involves negligible costs, and even more substantial curatorial choices fall within activities that museums routinely undertake. These features make targeted cultural framings a scalable and low-cost complement to traditional tools aimed at fostering civic and prosocial attitudes.

In this sense, museums and cultural institutions can operate as low-cost environments that temporarily increase the salience of socially valuable causes precisely when individuals are asked to act. This interpretation is consistent with our behavioral evidence and does not require the intervention to permanently reshape stated political or social attitudes.

\section{Conclusions}\label{sec:conclusion}

This paper investigates whether cultural institutions can be used as effective tools to promote prosocial behavior. We conducted a randomized controlled trial inside the \emph{Santa Maria della Scala} museum in Siena, introducing a treatment that emphasizes the site’s historical role in welcoming and caring for foreigners. The treatment led to a significant increase in charitable donations toward an NGO supporting refugees, and it was entirely driven by women. This is the outcome the intervention was designed to produce, showing that when cultural institutions are used with a clear purpose, they can effectively promote targeted prosocial behaviors.

In our setting, the emphasis on hospitality—a theme deeply rooted in the museum’s civic history—appears to activate behavioral responses consistent with increased generosity. Our paper aligns with a broader literature showing that prosocial behavior is sensitive to contextual cues, social image, and moral salience \citep{ariely2009doing,dellavigna2012testing}. It also resonates with models in which attention and situational framing alter public-good contributions \citep{bordalo2013salience,kessler2018identity}, and where social preferences are shaped by institutional or cultural context \citep{benabou2006incentives,fehr2011introduction}.

Taken together, our findings contribute to the growing literature on behavioral public economics by showing how culturally grounded nudges can shape prosocial decisions. They point to a broader role for institutions of memory, not only as preservers of the past, but as catalysts for present-day public good provision.

\setlength{\bibhang}{0pt}
\bibliographystyle{apalike}
\bibliography{bibliography.bib}

\begin{appendices}

\section{Self-reported measures of attitudes toward migration from ESS}\label{secA_ESS_mig}

Questions on welcoming of different cultures: for all of them subjects have to pick one of the following answers: {\emph i} Allow many to come and live here (4 points), {\emph{ii}} Allow some (3 points) {\emph {iii}} Allow a few (2 points), {\emph {iv}} Allow none (1 point).

\begin{itemize}
    \item \textbf{B29} To what extent do you think Italy should allow people of the same race or ethnic group as most Italy’s people to come and live here?
    \item \textbf{B30} To what extent do you think Italy should allow people of a different race or ethnic group of most Italian people to come and live here? 
    \item \textbf{B30a} And how about people from the poorer countries in Europe? Still use this card.
    \item \textbf{B31} How about people from the poorer countries outside Europe? Use the same card.
\end{itemize}

Questions on intercultural contact/exchange: for each of them, answer from 00 to 10 according to how good do they think it is the scenario proposed in the question.

\begin{itemize}
    \item \textbf{B32} Would you say it is generally bad or good for Italy’s economy that people come to live here from other countries? Please use this card.
    \item \textbf{B33} And, using this card, would you say that Italy’s cultural life is generally undermined or enriched by people coming to live here from other countries?
    \item \textbf{B34} Is Italy made a worse or a better place to live by people coming to live here from other countries?
\end{itemize}

\setcounter{table}{0}
\renewcommand{\thetable}{B\arabic{table}}

\setcounter{figure}{0}
\renewcommand{\thefigure}{B\arabic{figure}}

\section{Calendar}\label{app:calendar}

The tours were designed and prepared in collaboration with the museum guides, Deborah Barbagli and Nora Giordano, during September 2024 and January 2025. In March-April 2025, we set up the tour calendar jointly with them. Tours were scheduled on Mondays, Wednesdays, and Fridays, with each date randomly assigned to either the control or treatment condition, subject to the guides’ availability. The detailed allocation of control and treatment tours is reported below.

\begin{table}[ht!]
\centering
\caption{Calendar of museum tours used in the experiment, with allocation to control and treatment groups.}
\label{tab:calendar}
\begin{tabular}{lll}
\hline
Date & Day & Group \\
\hline
\multicolumn{3}{c}{March 2025} \\
\hline
3  & Monday    & Control \\
5  & Wednesday & Treatment \\
7  & Friday    & Control \\
10 & Monday    & Control \\
12 & Wednesday & Treatment \\
14 & Friday    & Treatment \\
17 & Monday    & Treatment \\
19 & Wednesday & Control \\
21 & Friday    & Control \\
24 & Monday    & Control \\
26 & Wednesday & Treatment \\
28 & Friday    & Treatment \\
31 & Monday    & Control \\
\hline
\multicolumn{3}{c}{April 2025} \\
\hline
2  & Wednesday & Treatment \\
7  & Monday    & Treatment \\
9  & Wednesday & Control \\
\hline
\end{tabular}
\end{table}

\setcounter{table}{0}
\renewcommand{\thetable}{C\arabic{table}}

\setcounter{figure}{0}
\renewcommand{\thefigure}{C\arabic{figure}}

\section{Additional Figures and Tables}\label{app:tables}

\subsection{Balance Table}

The treatment and control groups are balanced on gender, age, nationality, and immigration attitudes. However, enjoyment ratings are significantly higher in the control group ($p < 0.01$). Despite this, participants in the treatment group donated more, further ruling out satisfaction as a driver of the treatment effect. This interpretation is reinforced by the fact that enjoyment does not significantly predict donations in any model specification.

\input{balance}




\subsection{Donation Probabilities - Logit}

The logit regressions in Table \ref{tab:logit} yield results consistent with the probit estimates reported in the main text.

\input{logit}

\subsection{Male Sample Analysis}

Tables~\ref{tab:male_prob} and~\ref{tab:male_tobit} report Tobit and probit regressions estimated for male participants only, confirming the absence of statistically significant treatment effects.

\input{male_probability}

\input{male_tobit}

\clearpage

\subsection{Immigration Attitudes}
Table \ref{tab:imm_attitudes} shows that the treatment has no statistically significant effect on immigration attitudes ($\beta = -0.0166$, $p = 0.423$), confirming that the intervention did not shift participants' stated views on migration.
\input{reg_imm_att}

\paragraph{Mediation Analysis.}  Table \ref{tab:tobit_mediated} reports a Tobit regression of donation amounts on treatment and immigration attitudes. Notably, both the treatment effect ($\beta = 1.800$, $p < 0.05$) and the effect of immigration attitudes ($\beta = 5.204$, $p < 0.05$) are positive and statistically significant. Importantly, the treatment coefficient remains virtually unchanged when controlling for immigration attitudes (compared to $\beta = 1.769$ in the baseline model without attitudes), indicating that the increase in donations is not mediated by attitudinal change. 

\input{tobit_mediated}

\clearpage

\setcounter{table}{0}
\renewcommand{\thetable}{D\arabic{table}}

\setcounter{figure}{0}
\renewcommand{\thefigure}{D\arabic{figure}}

\section{Outliers}\label{app:outliers}

We examine how the results are influenced by the three outliers excluded from the main analysis. All three share very similar socio-demographic and behavioral characteristics: they are Italian females, they donated the full endowment split equally between the two NGOs, they reported the maximum enjoyment score (10), and they display a very high immigration attitude index. Notably, two of the three were assigned to the control group, and their donations account for 20\% of the total amount donated by all participants in that group.

\input{outliers_stat}

We consider it appropriate to exclude these outliers for several reasons. First, their presence heavily distorts the distribution of donations, particularly in the control group (see Table \ref{tab:outliers_stat}). Second, it is reasonable to assume that had the two control outliers been assigned to the treatment group, they would still have donated the full amount. For these reasons, we believe it is more appropriate to remove the outliers—even though no exclusion criterion was pre-registered—rather than allow the results of the experiment to be driven by the behavior of only three participants.

Here we report two additional robustness checks addressing the role of extreme donations. Both robustness checks retain the full sample, including the three full-endowment donations. In Table \ref{tab:wind}, we reduce the influence of extreme values by winsorizing donation amounts at the 95th percentile. In Table \ref{tab:log}, we instead estimate OLS specifications using the log-transformed outcome, defined as $\log(1+\text{donation})$, which preserves zero donations while compressing large positive donations.

The results are broadly consistent with the main findings, although statistical significance is weaker under these more conservative specifications. In the winsorized Tobit models, the treatment effect remains significant for total donations and CIR donations, but only at the 10 percent level in the full sample; among female participants, both effects remain significant at the 5 percent level. In the log-donation specifications, the effect on total donations is no longer significant at conventional levels, with p-values of 0.137 in the full sample and 0.103 among female participants. The effect on CIR donations remains marginally significant, while no effect is detected for LIPU. Overall, these checks confirm the same pattern as the main analysis, but show that the statistical strength of the result is somewhat sensitive to how extreme donations are handled.

\input{wind_tobit}
\input{log_tobit}
\end{appendices}

\end{document}

%% file: probit.tex
\begin{table}[htbp]\centering
\def\sym#1{\ifmmode^{#1}\else\(^{#1}\)\fi}
\caption{This table reports average marginal effects from probit regressions on the probability of donating. The first three columns use the full sample, while the last three columns restrict the sample to female participants. The columns report results for the probability of donating to any NGO, to CIR, and to LIPU. Full-sample specifications control for gender, age, and nationality (Italian); female-sample specifications control for age and nationality. Estimates for male participants are not statistically significant and are reported in the appendix.}
\label{tab:probit}
\resizebox{\textwidth}{!}{%
\begin{tabular}{l*{6}{c}}
\toprule
                &\multicolumn{1}{c}{All NGOs}&\multicolumn{1}{c}{CIR}&\multicolumn{1}{c}{LIPU}&\multicolumn{1}{c}{All NGOs (F)}&\multicolumn{1}{c}{CIR (F)}&\multicolumn{1}{c}{LIPU (F)}\\
\midrule
Treatment       &    0.069\sym{*}  &    0.054\sym{**} &    0.018         &    0.101\sym{**} &    0.073\sym{*}  &    0.008         \\
                &  (0.038)         &  (0.026)         &  (0.033)         &  (0.050)         &  (0.039)         &  (0.034)         \\
\addlinespace
Female          &    0.077\sym{**} &    0.106\sym{***}&    0.008         &                  &                  &                  \\
                &  (0.039)         &  (0.034)         &  (0.033)         &                  &                  &                  \\
\addlinespace
Age             &    0.015\sym{*}  &    0.010         &    0.009         &    0.023\sym{**} &    0.016         &    0.018\sym{***}\\
                &  (0.009)         &  (0.008)         &  (0.006)         &  (0.010)         &  (0.010)         &  (0.006)         \\
\addlinespace
Italian         &   -0.175         &   -0.138         &   -0.156         &   -0.237         &   -0.251\sym{*}  &   -0.188\sym{*}  \\
                &  (0.126)         &  (0.110)         &  (0.095)         &  (0.153)         &  (0.137)         &  (0.112)         \\
\midrule
Observations    &      384         &      384         &      384         &      247         &      247         &      247         \\
\bottomrule
\multicolumn{7}{l}{\footnotesize Standard errors in parentheses clustered at session level}\\
\multicolumn{7}{l}{\footnotesize \sym{*} \(p<0.10\), \sym{**} \(p<0.05\), \sym{***} \(p<0.01\)}\\
\end{tabular}
}
\end{table}

%% file: tobit.tex
\begin{table}[htbp]\centering
\def\sym#1{\ifmmode^{#1}\else\(^{#1}\)\fi}
\caption{This table reports Tobit regressions on donation amounts. The first column shows results for the total donations to both NGOs (CIR and LIPU), the second for donations to CIR, and the third for donations to LIPU. All specifications control for gender (female), age and nationality (Italian).}
\label{tab:tobit}
\resizebox{\textwidth}{!}{%
\begin{tabular}{l*{6}{c}}
\toprule
                &\multicolumn{1}{c}{All NGOs}&\multicolumn{1}{c}{CIR}&\multicolumn{1}{c}{LIPU}&\multicolumn{1}{c}{All NGOs (F)}&\multicolumn{1}{c}{CIR (F)}&\multicolumn{1}{c}{LIPU (F)}\\
\midrule
Treatment       &    1.769\sym{**} &    1.759\sym{**} &    0.883         &    2.479\sym{**} &    2.182\sym{**} &    0.842         \\
                &  (0.832)         &  (0.801)         &  (0.973)         &  (1.095)         &  (0.871)         &  (1.292)         \\
\addlinespace
Female          &    1.923\sym{**} &    2.789\sym{***}&    0.119         &                  &                  &                  \\
                &  (0.852)         &  (0.835)         &  (1.125)         &                  &                  &                  \\
\addlinespace
Age             &    0.393\sym{**} &    0.291         &    0.337\sym{**} &    0.544\sym{**} &    0.377         &    0.682\sym{***}\\
                &  (0.183)         &  (0.210)         &  (0.154)         &  (0.215)         &  (0.230)         &  (0.135)         \\
\addlinespace
Italian         &   -3.555         &   -2.979         &   -4.469\sym{**} &   -4.644\sym{*}  &   -4.702\sym{**} &   -5.798\sym{**} \\
                &  (2.338)         &  (2.386)         &  (2.159)         &  (2.806)         &  (2.356)         &  (2.783)         \\
\midrule
Observations    &      381         &      381         &      381         &      244         &      244         &      244         \\
\bottomrule
\multicolumn{7}{l}{\footnotesize Standard errors in parentheses clustered at session level}\\
\multicolumn{7}{l}{\footnotesize \sym{*} \(p<0.10\), \sym{**} \(p<0.05\), \sym{***} \(p<0.01\)}\\
\end{tabular}
}
\end{table}

%% file: ttest_gender.tex
\begin{table}[htbp]
\centering
\caption{This table reports mean donation outcomes by gender for the Control and Treatment groups, including total donations, likelihood of donating, and donations to each NGO (CIR and LIPU). For each measure, the difference between male and female participants, its p-value, and the number of observations are shown.}
\label{tab:ttest_gender}
\resizebox{\textwidth}{!}{%
\begin{tabular}{l*{5}{c} *{5}{c}}
\toprule
& \multicolumn{5}{c}{Control} & \multicolumn{5}{c}{Treatment} \\
\cmidrule(lr){2-6} \cmidrule(lr){7-11}
& Male & Female & Difference & p-value & Obs. & Male & Female & Difference & p-value & Obs. \\
\midrule
Amount Donated & 0.68 & 1.19 & -0.51 & 0.233 & 191      & 0.67&     1.65&    -0.98\sym{**} &    0.016&      195 \\
\addlinespace
Has Donated &  0.18&     0.19&    -0.01         &    0.865&      191      & 0.17&     0.31&    -0.14\sym{**} &    0.035&      195 \\
\addlinespace
Amount Donated to CIR & 0.45&     0.81&    -0.36         &    0.218&      191       &   0.47&     1.21&    -0.74\sym{**} &    0.025&      195 \\
\addlinespace
Amount Donated to LIPU &  0.23&     0.38&    -0.15         &    0.456&      191       &0.20&     0.44&    -0.25         &    0.203&      195 \\
\addlinespace
Has Donated to CIR & 0.11&     0.17&    -0.06         &    0.252&      191      &0.11&     0.25&    -0.14\sym{**} &    0.021&      195 \\
\addlinespace
Has Donated to LIPU & 0.09&     0.10&    -0.01         &    0.909&      191      & 0.10&     0.11&    -0.01         &    0.758&      195 \\
\bottomrule
\multicolumn{4}{l}{\footnotesize \sym{*} \(p<0.10\), \sym{**} \(p<0.05\), \sym{***} \(p<0.01\)}\\
\end{tabular}%
}
\end{table}

%% file: balance.tex
{
\begin{table}[htbp]\centering
\def\sym#1{\ifmmode^{#1}\else\(^{#1}\)\fi}
\caption{This table reports summary statistics for participants in the control and treatment groups. The last column reports differences in means between groups. Standard deviations are reported in parentheses.}
\label{tab:balance}
\begin{tabular}{l*{3}{c}}
\hline
                    &\multicolumn{1}{c}{Control}&\multicolumn{1}{c}{Treatment}&\multicolumn{1}{c}{Difference}\\
\hline
Female              &       0.65 &       0.64 &        0.02   \\
                    &     (0.48) &     (0.48) &      (0.05)   \\
Age                 &      23.72 &      23.84 &       -0.12   \\
                    &     (2.98) &     (2.67) &      (0.29)   \\
Italian             &       0.96 &       0.97 &       -0.01   \\
                    &     (0.19) &     (0.16) &      (0.02)   \\
Enjoyment           &       9.47 &       8.63 &        0.84***\\
                    &     (1.13) &     (1.70) &      (0.15)   \\
Immigration Attitude&       0.78 &       0.77 &        0.01   \\
                    &     (0.17) &     (0.19) &      (0.02)   \\                    
\hline
N                   &         192&         196&         388   \\
\hline
\end{tabular}
\end{table}
}

%% file: logit.tex
\begin{table}[htbp]\centering
\def\sym#1{\ifmmode^{#1}\else\(^{#1}\)\fi}
\caption{This table reports average marginal effects from logit regressions on the probability of donating. The first three columns use the full sample, while the last three columns restrict the sample to female participants. The columns report results for the probability of donating to any NGO, to CIR, and to LIPU. Full-sample specifications control for gender, age, and nationality (Italian); female-sample specifications control for age and nationality.}
\label{tab:logit}
\resizebox{\textwidth}{!}{%
\begin{tabular}{l*{6}{c}}
\toprule
                &\multicolumn{1}{c}{All NGOs}&\multicolumn{1}{c}{CIR}&\multicolumn{1}{c}{LIPU}&\multicolumn{1}{c}{All NGOs (F)}&\multicolumn{1}{c}{CIR (F)}&\multicolumn{1}{c}{LIPU (F)}\\
\midrule
Treatment       &    0.070\sym{*}  &    0.055\sym{**} &    0.017         &    0.100\sym{**} &    0.072\sym{*}  &    0.005         \\
                &  (0.039)         &  (0.027)         &  (0.034)         &  (0.050)         &  (0.039)         &  (0.036)         \\
\addlinespace
Female          &    0.081\sym{**} &    0.111\sym{***}&    0.012         &                  &                  &                  \\
                &  (0.039)         &  (0.036)         &  (0.034)         &                  &                  &                  \\
\addlinespace
Age             &    0.015\sym{*}  &    0.010         &    0.009\sym{*}  &    0.023\sym{**} &    0.016\sym{*}  &    0.018\sym{***}\\
                &  (0.008)         &  (0.008)         &  (0.005)         &  (0.010)         &  (0.009)         &  (0.006)         \\
\addlinespace
Italian         &   -0.168         &   -0.133         &   -0.140\sym{*}  &   -0.230         &   -0.239\sym{*}  &   -0.175\sym{*}  \\
                &  (0.117)         &  (0.099)         &  (0.081)         &  (0.145)         &  (0.126)         &  (0.095)         \\
\midrule
Observations    &      384         &      384         &      384         &      247         &      247         &      247         \\
\bottomrule
\multicolumn{7}{l}{\footnotesize Standard errors in parentheses clustered at session level}\\
\multicolumn{7}{l}{\footnotesize \sym{*} \(p<0.10\), \sym{**} \(p<0.05\), \sym{***} \(p<0.01\)}\\
\end{tabular}
}
\end{table}

%% file: male_probability.tex
\begin{table}[htbp]\centering
\def\sym#1{\ifmmode^{#1}\else\(^{#1}\)\fi}
\caption{This table reports average marginal effects from probit and logit regressions estimated for male participants only. The first three columns report probit estimates, while the last three report logit estimates. The dependent variables are indicators for donating to any NGO, to CIR, and to LIPU. All specifications control for age and nationality (Italian).}
\label{tab:male_prob}
\resizebox{\textwidth}{!}{%
\begin{tabular}{l*{6}{c}}
\toprule
                &\multicolumn{1}{c}{All NGOs (P)}&\multicolumn{1}{c}{CIR (P)}&\multicolumn{1}{c}{LIPU (P)}&\multicolumn{1}{c}{All NGOs (L)}&\multicolumn{1}{c}{CIR (L)}&\multicolumn{1}{c}{LIPU (L)}\\
\midrule
Treatment       &   -0.007         &    0.007         &    0.010         &   -0.008         &    0.007         &    0.008         \\
                &  (0.050)         &  (0.036)         &  (0.061)         &  (0.050)         &  (0.036)         &  (0.061)         \\
\addlinespace
Age             &    0.003         &    0.004         &   -0.004         &    0.003         &    0.004         &   -0.003         \\
                &  (0.010)         &  (0.007)         &  (0.009)         &  (0.010)         &  (0.006)         &  (0.008)         \\
\addlinespace
Italian         &   -0.109         &    0.000         &   -0.186         &   -0.105         &    0.000         &   -0.161         \\
                &  (0.200)         &      (.)         &  (0.134)         &  (0.177)         &      (.)         &  (0.113)         \\
\midrule
Observations    &      137         &      134         &      137         &      137         &      134         &      137         \\
\bottomrule
\multicolumn{7}{l}{\footnotesize Standard errors in parentheses clustered at session level}\\
\multicolumn{7}{l}{\footnotesize \sym{*} \(p<0.10\), \sym{**} \(p<0.05\), \sym{***} \(p<0.01\)}\\
\end{tabular}
}
\end{table}

%% file: male_tobit.tex
\begin{table}[htbp]\centering
\def\sym#1{\ifmmode^{#1}\else\(^{#1}\)\fi}
\caption{This table reports Tobit regressions on donation amounts. The first column shows results for the total donations to both NGOs (CIR and LIPU), the second for donations to CIR, and the third for donations to LIPU. All specifications control for age, nationality (Italian).}
\label{tab:male_tobit}
\begin{tabular}{l*{3}{c}}
\toprule
                &\multicolumn{1}{c}{All NGOs}&\multicolumn{1}{c}{CIR}&\multicolumn{1}{c}{LIPU}\\
\midrule   
Treatment       &   -0.025         &    0.301         &    0.177         \\
                &  (1.109)         &  (1.441)         &  (1.477)         \\
\addlinespace
Age             &    0.122         &    0.214         &   -0.068         \\
                &  (0.243)         &  (0.279)         &  (0.201)         \\
\addlinespace
Italian         &   -2.557         &   36.927\sym{***}&   -5.324         \\
                &  (4.620)         &  (5.870)         &  (3.542)         \\
\midrule
Observations    &      137         &      137         &      137         \\
\bottomrule
\multicolumn{4}{l}{\footnotesize Standard errors in parentheses clustered at session level}\\
\multicolumn{4}{l}{\footnotesize \sym{*} \(p<0.10\), \sym{**} \(p<0.05\), \sym{***} \(p<0.01\)}\\
\end{tabular}
\end{table}

%% file: reg_imm_att.tex
\begin{table}[htbp]\centering
\def\sym#1{\ifmmode^{#1}\else\(^{#1}\)\fi}
\caption{This table reports OLS estimates of the effect of the treatment on the ESS-based migration attitude index. Controls include gender, age, and nationality (Italian).}
\label{tab:imm_attitudes}
\begin{tabular}{l*{1}{c}}
\toprule
                &\multicolumn{1}{c}{OLS}\\
\midrule
Treatment       &  -0.0166         \\
                & (0.0202)         \\
\addlinespace
Female          &   0.0674\sym{**} \\
                & (0.0242)         \\
\addlinespace
Age             &  0.00664\sym{**} \\
                &(0.00260)         \\
\addlinespace
Italian         &   0.0766\sym{*}  \\
                & (0.0398)         \\
\bottomrule
\multicolumn{2}{l}{\footnotesize Standard errors in parentheses}\\
\multicolumn{2}{l}{\footnotesize \sym{*} \(p<0.10\), \sym{**} \(p<0.05\), \sym{***} \(p<0.01\)}\\
\end{tabular}
\end{table}

%% file: tobit_mediated.tex
\begin{table}[htbp]\centering
\def\sym#1{\ifmmode^{#1}\else\(^{#1}\)\fi}
\caption{This table reports Tobit estimates of the effect of the treatment on the amount donated to NGOs, controlling for the ESS-based migration attitude index, gender, age, and nationality (Italian).}
\label{tab:tobit_mediated}
\begin{tabular}{l*{1}{c}}
\toprule
                &\multicolumn{1}{c}{All NGOs}\\
\midrule
Treatment       &    1.800\sym{**} \\
                &  (0.834)         \\
\addlinespace
Attitude Immigration&    5.204\sym{**} \\
                &  (2.377)         \\
\addlinespace
Female          &    1.546\sym{*}  \\
                &  (0.800)         \\
\addlinespace
Age             &    0.360\sym{**} \\
                &  (0.181)         \\
\addlinespace
Italian         &   -3.919         \\
                &  (2.429)         \\
\bottomrule
\multicolumn{2}{l}{\footnotesize Standard errors in parentheses}\\
\multicolumn{2}{l}{\footnotesize \sym{*} \(p<0.10\), \sym{**} \(p<0.05\), \sym{***} \(p<0.01\)}\\
\end{tabular}
\end{table}

%% file: outliers_stat.tex
\begin{table}[htbp]
\centering
\caption{Summary statistics of donations by experimental group, with and without the three outliers. Reported statistics include the number of observations (N), mean, standard deviation (sd), skewness, and kurtosis.}
\label{tab:outliers_stat}
\resizebox{0.8\textwidth}{!}{%
\def\sym#1{\ifmmode^{#1}\else\(^{#1}\)\fi}
\begin{tabular}{l*{1}{ccccc}}
\toprule
            &       N&        mean&          sd&    skewness&    kurtosis\\
\hline
Full sample (control)   &         192&    1.04&    2.81&    4.13&    24.32\\
No outliers (control)   &         190&    0.84&    2.04&    2.71&    10.41\\
Full sample (treatment)   &         196&    1.29&    2.76&     2.98&    15.01\\
No outliers (treatment)   &         195&    1.19&    2.41&    2.18&    7.28\\
\bottomrule
\end{tabular}
}
\end{table}

%% file: wind_tobit.tex
\begin{table}[htbp]\centering
\def\sym#1{\ifmmode^{#1}\else\(^{#1}\)\fi}
\caption{This table reports Tobit regressions on winsorized donation amounts, with donations winsorized at the 95th percentile. The first three columns use the full sample, while the last three columns restrict the sample to female participants. The columns report results for total donations to both NGOs, donations to CIR, and donations to LIPU. Full-sample specifications control for gender, age, and nationality (Italian); female-sample specifications control for age and nationality.}
\label{tab:wind}
\resizebox{\textwidth}{!}{%
\begin{tabular}{l*{6}{c}}
\toprule
                &\multicolumn{1}{c}{All NGOs}&\multicolumn{1}{c}{CIR}&\multicolumn{1}{c}{LIPU}&\multicolumn{1}{c}{All NGOs (F)}&\multicolumn{1}{c}{CIR (F)}&\multicolumn{1}{c}{LIPU (F)}\\
\midrule
Treatment       &    1.598\sym{*}  &    1.548\sym{*}  &    0.635         &    2.195\sym{**} &    1.877\sym{**} &    0.445         \\
                &  (0.843)         &  (0.794)         &  (0.956)         &  (1.019)         &  (0.836)         &  (1.084)         \\
\addlinespace
Female          &    2.121\sym{***}&    3.027\sym{***}&    0.462         &                  &                  &                  \\
                &  (0.770)         &  (0.791)         &  (1.004)         &                  &                  &                  \\
\addlinespace
Age             &    0.380\sym{**} &    0.279         &    0.312\sym{*}  &    0.519\sym{**} &    0.358         &    0.600\sym{***}\\
                &  (0.191)         &  (0.212)         &  (0.163)         &  (0.215)         &  (0.231)         &  (0.139)         \\
\addlinespace
Italian         &   -3.357         &   -2.712         &   -4.091\sym{**} &   -4.322         &   -4.341\sym{**} &   -5.014\sym{**} \\
                &  (2.239)         &  (2.236)         &  (2.003)         &  (2.653)         &  (2.194)         &  (2.496)         \\
\midrule
Observations    &      384         &      384         &      384         &      247         &      247         &      247         \\
\bottomrule
\multicolumn{7}{l}{\footnotesize Standard errors in parentheses clustered at session level}\\
\multicolumn{7}{l}{\footnotesize \sym{*} \(p<0.10\), \sym{**} \(p<0.05\), \sym{***} \(p<0.01\)}\\
\end{tabular}
}
\end{table}

%% file: log_tobit.tex
\begin{table}[htbp]\centering
\def\sym#1{\ifmmode^{#1}\else\(^{#1}\)\fi}
\caption{This table reports OLS regressions on log-transformed donation amounts, defined as $\log(1+\text{donation})$. The first three columns use the full sample, while the last three columns restrict the sample to female participants. The columns report results for total donations to both NGOs, donations to CIR, and donations to LIPU. Full-sample specifications control for gender, age, and nationality (Italian); female-sample specifications control for age and nationality.}
\label{tab:log}
\resizebox{\textwidth}{!}{%
\begin{tabular}{l*{6}{c}}
\toprule
                &\multicolumn{1}{c}{All NGOs}&\multicolumn{1}{c}{CIR}&\multicolumn{1}{c}{LIPU}&\multicolumn{1}{c}{All NGOs (F)}&\multicolumn{1}{c}{CIR (F)}&\multicolumn{1}{c}{LIPU (F)}\\
\midrule
Treatment       &    0.115         &    0.092\sym{*}  &    0.015         &    0.162         &    0.127\sym{*}  &    0.008         \\
                &  (0.073)         &  (0.051)         &  (0.047)         &  (0.093)         &  (0.066)         &  (0.051)         \\
\addlinespace
Female          &    0.183\sym{***}&    0.180\sym{***}&    0.044         &                  &                  &                  \\
                &  (0.054)         &  (0.045)         &  (0.039)         &                  &                  &                  \\
\addlinespace
Age             &    0.032\sym{*}  &    0.018         &    0.018\sym{*}  &    0.046\sym{**} &    0.026         &    0.029\sym{**} \\
                &  (0.017)         &  (0.014)         &  (0.010)         &  (0.021)         &  (0.019)         &  (0.010)         \\
\addlinespace
Italian         &   -0.340         &   -0.158         &   -0.290         &   -0.426         &   -0.330         &   -0.265         \\
                &  (0.265)         &  (0.177)         &  (0.225)         &  (0.295)         &  (0.194)         &  (0.241)         \\
\midrule
Observations    &      384         &      384         &      384         &      247         &      247         &      247         \\
\bottomrule
\multicolumn{7}{l}{\footnotesize Standard errors in parentheses clustered at session level}\\
\multicolumn{7}{l}{\footnotesize \sym{*} \(p<0.10\), \sym{**} \(p<0.05\), \sym{***} \(p<0.01\)}\\
\end{tabular}
}
\end{table}